\documentclass[12pt]{article}

\usepackage{graphicx}
\input epsf
\usepackage[usenames]{color}

\begin{document}

\begin{titlepage}
\begin{flushright}
CALT-68-2581\\
ITEP-TH-69/05
\end{flushright}

\begin{center}
{\Large\bf $ $ \\ $ $ \\
A nonlocal Poisson bracket of the sine-Gordon model 
}\\
\bigskip\bigskip\bigskip
{\large Andrei Mikhailov}
\\
\bigskip\bigskip
{\it California Institute of Technology 452-48,
Pasadena CA 91125 \\
\bigskip
and\\
\bigskip
Institute for Theoretical and 
Experimental Physics, \\
117259, Bol. Cheremushkinskaya, 25, 
Moscow, Russia}\\

\vskip 1cm
\end{center}

\begin{abstract}
	It is well known that the classical string on a two-sphere
	is more or less equivalent to	the sine-Gordon model. 
	We consider the nonabelian 
	dual of the classical string on a two-sphere. We show that
	there is a projection map
	from the phase space of this model to the phase space of
	the sine-Gordon model. The corresponding Poisson structure of
	the sine-Gordon model is nonlocal with one integration.
\end{abstract}

\end{titlepage}

\section{Introduction.}
The most well-known example of the AdS/CFT correspondence 
is the duality between the
Type IIB superstring in $AdS_5\times S^5$ and
the ${\cal N}=4$ supersymmetric Yang-Mills theory on
${\bf R}\times S^3$. At the level of the classical string we
can consider a simpler example when the motion of the string is 
restricted to ${\bf R}\times S^2\subset AdS_5\times S^5$.
The classical string on ${\bf R}\times S^2$
is essentially equivalent to a well-known integrable system,
the sine-Gordon model \cite{Pohlmeyer:1975nb}. 
But the symplectic structure
of the classical string does not correspond to the canonical
symplectic structure of the sine-Gordon model. 

In this paper we will argue that the map from the classical
string to the sine-Gordon can be understood as a kind of
T-duality in $S^2$.
We consider the infinite string on ${\bf R}\times S^2$. 
In Section \ref{sec:TDual} we introduce the classical field
theory which is dual to the classical string on ${\bf R}\times S^2$.
After imposing the Virasoro constraints the phase space
becomes an affine bundle over the phase space of the sine-Gordon,
modelled on the vector bundle of solutions of some auxiliary linear
problem. The distribution of symplectic complements of the fibers is
integrable. In Section \ref{sec:Poisson} we  explain
how to restrict the symplectic form of the string
to the integral manifold of this distribution and push it forward
to the sine-Gordon phase space. This gives a nonlocal Poisson bracket
of the sine-Gordon which is compatible with the standard Poisson bracket
obtained from the sine-Gordon action. 
In Section \ref{sec:UsualString} we return to the original
system, the classical string on a sphere, and 
show that it leads to  the same Poisson structure
of the  sine-Gordon model.
This Poisson bracket and the corresponding symplectic form
are given by Eqs. (\ref{OmegaOnL}), (\ref{Compatibility}) and
(\ref{NonlocalPB}).

This suggests that the quantization of the sigma-model on 
${\bf R}\times S^2$ (after imposing the Virasoro constraints) could
be closely related to the quantization of the sine-Gordon model
with the nonlocal Poisson bracket. 

A somewhat similar relation between the third Poisson structure of KdV
and the WZNW model was obtained in \cite{Gorsky:1994ee}.
The general theory of nonlocal Poisson brackets was developed in
\cite{MaltsevNovikov} and references therein.
The importance of the non-standard Poisson brackets for
 AdS/CFT was emphasized in \cite{Zarembo:2004hp}.
T-duality with respect to a nonabelian symmetry was discussed
in 
\cite{FridlingJevicki,FradkinTseytlin,delaOssaQuevedo,CurtrightZachos}.

{\em Note added in the revised version.}
Closely related results were previously obtained in 
\cite{Doliwa:1994bk}---\cite{MR2180896}, but from a different perspective.  
The main difference of our approach is that we start from the 
relativistic string and derive the Poisson brackets
from the string worldsheet action. 
(While in \cite{Doliwa:1994bk}---\cite{MR2180896} the Poisson brackets were essentially postulated.)
The results of this work are extended to the superstring in $AdS_5\times S^5$ 
in \cite{Bihamiltonian}.

 \section{T-dual of the classical string on a sphere.}\label{sec:TDual}
The zero-curvature approach to the string on a sphere was suggested
in the context of AdS/CFT in \cite{Bena:2003wd} and further
developed in \cite{Beisert:2005bm}. We start with the Lie
algebra $\bf g$ and its subalgebra $\bf h$. Suppose that as a linear space
${\bf g}={\bf h}+{\bf j}$ where ${\bf j}={\bf h}^{\perp}$.
Suppose that $[{\bf j},{\bf j}]\subset {\bf h}$.
Consider a pair of the Lie algebra valued fields $J_{\pm}, H_{\pm}$
such that $J_{\pm}\in {\bf j}$ and $H_{\pm}\in {\bf h}$. Consider
the following zero curvature equations:
\begin{equation}\label{ZeroCurvature}
	\left[\partial_+ + H_+ +{1\over z} J_+,
	\partial_- +H_- + z J_-\right]=0
\end{equation}
In particular case when ${\bf g}=so(n+1)$ and ${\bf h}=so(n)$
the coset space is a sphere $S^n$ and 
the fields $H_{\pm}$ and $J_{\pm}$ have a very transparent 
geometrical meaning. Let us choose a local trivialization
of the tangent bundle $TS^n$, that is specify at each point
$x\in S^n$ a basis $e_1(x),\ldots,e_n(x)$ in the tangent space.
Given the embedding of the classical string worldsheet
$x(\tau,\sigma)$ we write
\[
	\partial_{\pm} x = \sum_{j=1}^n \phi^j_{\pm} e_j
\]
We put
\[
	J_{\pm}=\left[
	\begin{array}{cccc}
		0 & \phi^1_{\pm} & \ldots & \phi^n_{\pm} \\
	-\phi^1_{\pm} & & & \\
	\vdots &  & 0&\\
	-\phi^n_{\pm} & & & 
\end{array}\right]
\]
and define $H_{\pm}$ as an antisymmetric $n\times n$ matrix 
$H_{\pm}^{ij}$ such that
\[
	D_{\pm} e_i =  e^j H_{\pm}^{ji}
\]
Then equations (\ref{ZeroCurvature}) encode the string equations of motion
\[
	D_+\partial_-x=D_-\partial_+x=0
\]
and the relation between the Riemann tensor and the metric tensor
for the sphere:
\[
	R^i_{jkl} \partial_+ x^k \partial_- x^l=
	\partial_+ x^i \partial_- x^j -\partial_- x^i \partial_+ x^j
\]
Let us now consider a particular case when $n=3$, the target space is $S^2$.
The point of the two-sphere is usually denoted ${\bf n}$: $x={\bf n}$.
We can think of ${\bf n}$ as a unit vector in ${\bf R}^3$.
Suppressing the vector index $i=1,2$ we can write:
\[
	\phi_{\pm}=\partial_{\pm}{\bf n}
\]
If $\xi$ is a section of the restriction to the string worldsheet of the
tangent bundle to the sphere, then we have
\begin{equation}
	[D_+,D_-]\xi=\phi_+(\phi_-,\xi)-\phi_-(\phi_+,\xi)
\end{equation}
Define the operator $I$ on the tangent space as a rotation
by $\pi\over 2$, so that $I^2=-1$. We will consider the classical
field theory with the following action:
\begin{eqnarray}\label{DualAction}
&&	S=\int d\tau d\sigma \left\{
	\left(\phi_+,(1-\psi I) \phi_-\right)
	+\right.\nonumber\\[5pt]
&&	\left.+(\phi_+,D_-\lambda)
	-(\phi_-,D_+\lambda)+
	\psi(\partial_+A_--\partial_- A_+)\right\}
	\label{Action}
\end{eqnarray}
where $\lambda$ and $\psi$ are Lagrange multipliers 
and 
\[ D_{\pm}=\partial_{\pm}+IA_{\pm} \]   
We conjecture that the quantum theory with the action (\ref{DualAction})
is equivalent to the string on $S^2$. We do not have a solid argument
for this, but naively 
if we integrate out $\lambda$ and $\psi$ we return to the standard
action $\int d\tau d\sigma (\phi_+,\phi_-)$ of the classical string 
on $S^2$. It is possible that the correct statement of quantum
equivalence would require the supersymmetric 
extension of the model\footnote{I want to thank A.~Tseytlin for
the correspondence on these things.}. 
We will now study the Poisson structure of the theory with the 
action (\ref{DualAction}) and then in Section \ref{sec:UsualString}
we will show that the standard action of the classical string
$\int d\tau^+d\tau^- (\partial_+ {\bf n},\partial_- {\bf n})$ 
leads to essentially
the same Poisson structure. 

The action (\ref{Action}) has a gauge symmetry corresponding to the
change of the basis in the tangent space to $S^2$:
\begin{equation}
	\delta\phi_{\pm}=\epsilon I.\phi_{\pm},\;\;
	\delta\lambda=\epsilon I.\lambda,\;\;
	\delta A_{\pm}=-\partial_{\pm}\epsilon,\;\;
	\delta \psi=0
\end{equation}
The equations of motion are:
\begin{eqnarray}
	&&	D_-\phi_+=D_+\phi_-=0\label{FirstGaugeInvariant}\\
&&	D_-\lambda=-(1-\psi I)\phi_-\\
&&	D_+\lambda=(1+\psi I)\phi_+\\
&&	\partial_+\psi=(\phi_+,I.\lambda)\\
&& 	\partial_-\psi=(\phi_-,I.\lambda)\\
&&	\partial_+ A_--\partial_- A_+=(\phi_+,I.\phi_-)
\label{LastGaugeInvariant}
\end{eqnarray}
The symplectic structure read from the action is:
\begin{equation}
	\omega=
	\oint \left\{ 
	\left[(\delta \phi_+, \delta\lambda)+
	 \delta A_+ \delta\psi\right]d\tau^+ +
	\left[(\delta \phi_-, \delta\lambda)
	+ \delta A_-\delta\psi\right]d\tau^-
	\right\}
\end{equation}
One can check that this symplectic structure is gauge-invariant
and does not depend on the choice of the contour on the worldsheet.
	
We will use a complex notation for two-dimensional vectors, for example
$\phi_{\pm}=\phi^1_{\pm}+i\phi^2_{\pm}$.
Let us denote $r_+=|\phi_+|$ and $r_-=|\phi_-|$.
Let us choose a special gauge:
\begin{equation}
	\mbox{Im}(\phi_+\phi_-)=0
\end{equation}
This means that
\[
\phi_{\pm}=r_{\pm}e^{\pm i\varphi}
\]
In this gauge the equations of motion are:
\begin{eqnarray}\label{EqMFirst}
&&	\partial_-r_+=\partial_+r_-=0\\
&&	A_+=\partial_+\varphi,\;\;\; A_-=-\partial_-\varphi\\
&&	\partial_-[e^{-i\varphi}\lambda]=
-r_-(1-i\psi)e^{-2i\varphi}\label{PartialMinusLambda}\\
&&	\partial_+[e^{i\varphi}\lambda] =
r_+(1+i\psi)e^{2i\varphi}\label{PartialPlusLambda}\\
&&	\partial_+\psi={ir_+\over 2}
 	(e^{-i\varphi}\lambda-e^{i\varphi}\overline{\lambda})
	\label{PartialPlusPsi}\\
&&	\partial_-\psi={ir_-\over 2}
	(e^{i\varphi}\lambda-e^{-i\varphi}\overline{\lambda})
	\label{PartialMinusPsi}\\
&&	\partial_+\partial_-\psi=
	-r_+r_-(\psi\cos 2\varphi +\sin 2\varphi)\\
&&	\partial_+\partial_-\varphi=-{1\over 2}r_+r_-\sin(2\varphi)
\label{EqMLast}
\end{eqnarray}
The symplectic form becomes:
\begin{eqnarray}
	\omega=\int\left\{
	d\tau^+\left[ 2\partial_+\delta\psi \delta\varphi +
	{1\over 2}\delta r_+ \delta (e^{-i\varphi}\lambda + 
	e^{i\varphi}\overline{\lambda})\right]-\right.\nonumber\\ \left.
	-d\tau^-\left[ 2\partial_-\delta\psi \delta\varphi +
	{1\over 2}\delta r_- \delta (e^{i\varphi}\lambda +
	e^{-i\varphi}\overline{\lambda})\right]\right\}
	\label{SymplecticForm}
\end{eqnarray}
For $r_+=r_-=1$ 
the field $\lambda$ satisfies the following differential equations:
\begin{eqnarray}
&&	\partial_+^2\lambda + \left[
(\partial_+\varphi)^2+i\partial_+^2\varphi 
+{1\over 2}\right]\lambda = {1\over 2}
e^{2i\varphi}\overline{\lambda}\label{AuxiliaryLambda}\\[5pt]
&&	\partial_-^2\lambda + \left[
(\partial_-\varphi)^2-i\partial_-^2\varphi 
+{1\over 2}\right]\lambda = {1\over 2}
e^{-2i\varphi}\overline{\lambda}
\end{eqnarray}
with the additional conditions:
\begin{equation}
	\partial_+(e^{i\varphi}\lambda)+
	\partial_-(e^{i\varphi}\overline{\lambda})=0
\end{equation}
\begin{equation}
	\mbox{Re}[e^{-2i\varphi}\partial_+(e^{i\varphi}\lambda)]=1
\end{equation}
Then $\psi$ is defined as:
\begin{equation}
	\psi=\mbox{Im}[e^{-2i\varphi}\partial_+(e^{i\varphi}\lambda)]
\end{equation}
The variations  $\Delta\lambda$ and $\Delta \psi$ at fixed $\varphi$
satisfy the auxiliary linear equations:
\begin{equation}\label{AuxiliaryLinear}
	\partial_+\left[\begin{array}{c}
		\Delta\lambda \\
		\overline{\Delta\lambda}\\
		\sqrt{2}\Delta\psi \end{array}
		\right]=
		\left[\begin{array}{ccc}
			-i\partial_+\varphi & 0 & ie^{i\varphi}/\sqrt{2}\\
			0 & i\partial_+\varphi & -ie^{-i\varphi}/\sqrt{2}\\
			ie^{-i\varphi}/\sqrt{2} & -ie^{i\varphi}/\sqrt{2} & 0
	\end{array}\right]
	\left[\begin{array}{c}
		\Delta\lambda \\
		\overline{\Delta\lambda}\\
		\sqrt{2}\Delta\psi 
	\end{array}
	\right]
\end{equation}
\begin{equation}\label{AuxiliaryLinearMinus}
	\partial_-\left[\begin{array}{c}
		\Delta\lambda \\
		\overline{\Delta\lambda}\\
		\sqrt{2}\Delta\psi \end{array}
		\right]=
		\left[\begin{array}{ccc}
			i\partial_-\varphi & 0 & ie^{-i\varphi}/\sqrt{2}\\
			0 & -i\partial_+\varphi & -ie^{i\varphi}/\sqrt{2}\\
			ie^{i\varphi}/\sqrt{2} & -ie^{-i\varphi}/\sqrt{2} & 0
	\end{array}\right]
	\left[\begin{array}{c}
		\Delta\lambda \\
		\overline{\Delta\lambda}\\
		\sqrt{2}\Delta\psi 
	\end{array}
	\right]
\end{equation}
Similar auxiliary linear 
equations  were considered
in \cite{NeveuPapanicolaou,Papanicolaou}.
Notice that $\psi$ can be expressed in terms of $\varphi$ from the equation:
\begin{equation}\label{LPsi}
	\partial_+ \left[
	{\partial_+^2 \psi + \psi\over q_+}\right] + 4 q_+ \partial_+\psi = 2
\end{equation}
where 
\[
	q_+=\partial_+\varphi
\]
This equation follows from (\ref{PartialMinusLambda})---(\ref{PartialMinusPsi}).
But it does not determine $\psi$ unambiguously because
the linear equation
\begin{equation}\label{LinearDeltaPsi}
	\partial_+ \left[
	{\partial_+^2 \Delta\psi + \Delta\psi\over q_+}\right] 
	+ 4 q_+ \partial_+\Delta\psi = 0
\end{equation}
has nontrivial solutions. There are three linearly independent solutions.
Therefore $\psi$ is determined by $q_+$ up to adding $\Delta \psi$ satisfying
Eq. (\ref{LinearDeltaPsi}). On the other hand, for each $\psi$ satisfying
(\ref{LPsi}) we can determine $\lambda$ unambiguously from 
Eqs. (\ref{PartialPlusPsi},\ref{PartialMinusPsi}). 
This tells us that the space of solutions $\varphi,\lambda,\psi$
of Eqs.
(\ref{EqMFirst})---(\ref{EqMLast}) for $r_+=r_-=1$
is an affine bundle\footnote{Affine bundle means that
the fibers are affine spaces. An affine space is almost a linear
space, but without $0$; we cannot add points, but we can consider
the ``difference'' between the two points. The difference
takes value in some linear space, on which the affine space
is ``modelled''.} over the space of solutions $\varphi$
of the sine-Gordon equation,  modelled on the vector bundle of
the solutions to the linear problem (\ref{AuxiliaryLinear}).
In other words, the space of solutions of the linear problem
(\ref{AuxiliaryLinear}) is precisely the ambiguity in 
restoring $\lambda$ and $\psi$ from $\varphi$. 
\begin{figure}
\begin{center} 
\epsfxsize=2.5in {\epsfbox{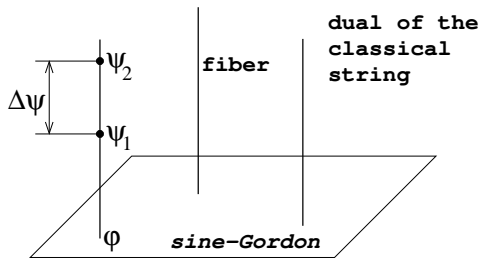}} 
\caption{The phase space of the classical string on an infinite line is an affine bundle
over the phase space of the sine-Gordon model.}
\end{center} 
\end{figure}

It is interesting to observe that each section $\Delta\psi$ of the
bundle of solutions of the auxiliary linear problem (\ref{AuxiliaryLinear})
defines a vector field on the sine-Gordon phase space: 
$\dot{\varphi}=\Delta\psi$. The map from the space of sections
of the ``auxiliary linear bundle'' to the vector fields on the sine-Gordon
phase space can be understood as follows. Take $\Delta\psi$, understand
it as a vector tangent to the fiber ${\cal F}$, then lower the index
by the symplectic structure (\ref{SymplecticForm}), then raise the index
by the standard Poisson structure of the sine-Gordon. 

It is also interesting that the vector field $\dot{\varphi}=\psi$
can be thought of as a variation of the sine-Gordon solution
with respect to the change of the mass parameter of the sine-Gordon (the 
coefficient in front of $\cos 2\varphi$ in the action).

\begin{figure}\label{fig:lc}
\begin{center} 
\epsfxsize=2.5in {\epsfbox{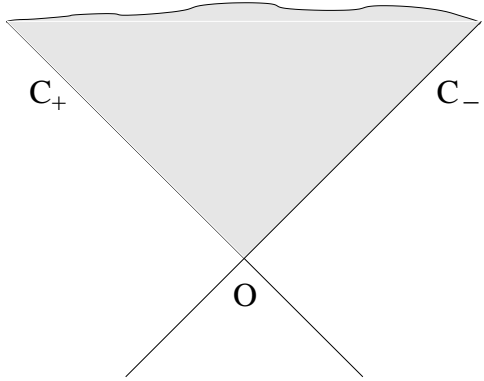}} 
\caption{The light cone}
\end{center} 
\end{figure}
In the rest of this paper we will use the ``light cone'' method for
describing the Poisson brackets. Let us
briefly explain this. Pick a point $O$ on the worldsheet with the
coordinates $(\tau_0^+,\tau_0^-)$ and consider the light cone
with the origin at this point, see Fig. \ref{fig:lc}. 
The light cone consists of two lines,
$C^+$ with $\tau^-=\tau^-_0$ and $C^-$ with $\tau^+=\tau^+_0$. 
The solution inside the shaded region is determined by the
data $(\varphi,\psi,\lambda)$ on $C^+_{\tau^+\geq 0}$ and
$C^-_{\tau^-\geq 0}$. Therefore we can describe the Poisson bracket
by saying what is the Poisson bracket of the fields on the light cone.
The standard Poisson bracket for $\varphi$ would be
\begin{eqnarray}
&&	\{ \varphi(\tau_1^+,\tau_0^-), \varphi(\tau_2^+,\tau_0^-)\}_{usual}=
	{1\over 2} \varepsilon(\tau_1^+-\tau_2^+)\nonumber\\
&&	\{\varphi(\tau_0^+,\tau_1^-), \varphi(\tau_0^+,\tau_2^-)\}_{usual}=
	{1\over 2} \varepsilon(\tau_1^--\tau_2^-)\nonumber
\end{eqnarray}
where $\varepsilon(\tau_1-\tau_2)$ is $1$ if $\tau_1>\tau_2$ and $-1$
if $\tau_1<\tau_2$.
But this is not the Poisson bracket corresponding to the 
symplectic form (\ref{SymplecticForm}).
We will now describe on the light cone the Poisson structure corresponding
to  the symplectic form (\ref{SymplecticForm}) with the constraint
$r_+=r_-=1$.

\section{Poisson structure.}
\label{sec:Poisson}
Let ${\cal M}_{\widetilde{O(3)}}$ denote the space of
solutions of (\ref{EqMFirst})---(\ref{EqMLast}) and 
${\cal M}_{\widetilde{CS}}$ denote the
space of solutions with $r_+=r_-=1$. (The index CS stands
for ``classical string'', and the tilde reminds us that
we are considering the T-dual model.) Let us denote ${\cal M}_{SG}$
the space of $\varphi$ solving the sine-Gordon equation.
We have seen that ${\cal M}_{\widetilde{CS}}$ is an affine bundle
over ${\cal M}_{SG}$. For a point $x\in {\cal M}_{\widetilde{CS}}$
let us denote ${\cal F}_x$ the fiber going through this point.
In other words, if $x=(\varphi,\lambda,\psi)$ then 
${\cal F}_x$ consists of all the solutions of the form
$(\varphi,\lambda+\Delta \lambda, \psi+\Delta\psi)$ where
$\Delta\lambda$ and $\Delta\psi$ satisfy (\ref{AuxiliaryLinear}).
Let $T_x{\cal F}_x\subset T_x{\cal M}_{\widetilde{CS}}$ be the tangent space 
to ${\cal F}_x$ at the point $x$. Let us denote 
$\widehat{\cal L}_x\subset T_x{\cal M}_{\widetilde{CS}}$ 
the subspace of the tangent
space to ${\cal M}_{\widetilde{CS}}$ 
at the point $x$ consisting of the vectors
orthogonal with respect to $\omega$ to $T_x{\cal F}_x$. In other
words, $\widehat{\cal L}_x$ is the space of all vectors 
$\xi\in T_x{\cal M}_{\widetilde{CS}}$
such that for any $\eta\in T_x{\cal F}_x$ we have $\omega(\xi,\eta)=0$.
Schematically: $\widehat{\cal L}_x=(T_x{\cal F}_x)^{\perp}$.
Let us denote ${\cal L}_{\varphi}=\widehat{\cal L}_x/T{\cal F}_x$.
We will view ${\cal L}_{\varphi}$
as a subspace in the tangent space to ${\cal M}_{SG}$.
We have a distribution of planes ${\cal L}_{\varphi}\subset 
T_{\varphi}{\cal M}_{SG}$.
We will now show that this distribution is integrable and defines
a foliation of ${\cal M}_{SG}$ of the codimension three.

Let us first introduce some notations. When we write a differential
operator in the space of functions of $\tau^+$ we understand that
each operator $\partial_+={\partial\over \partial \tau^+}$ acts on
everything to the right of it. For example:
\[
	\partial_+ f_1 f_2 = f_1 \partial_+ f_2 + f_2\partial_+ f_1
\]
But if a part of the expression is inside  the $\langle\rangle$
brackets, then any  $\partial_+$ inside the brackets acts only
on everything to the right of it inside the brackets, but not
to the right of the $\rangle$ bracket. For example:
\begin{eqnarray}
&&	\langle\partial_+ f_1 f_2\rangle f_3 =
	f_2 f_3\partial_+ f_1 + f_1 f_3 \partial_+ f_2\nonumber\\ 
&&	\partial_+ f_1 \langle\partial_+ f_2 \rangle f_3=
	f_3 \langle \partial_+ f_2 \rangle \partial_+ f_1+
	f_1 f_3 \partial_+^2 f_2 + 
	f_1\langle\partial_+ f_2\rangle \partial_+ f_3\nonumber
\end{eqnarray}
Let us denote by $L_+$ the operator:
\begin{equation}
	L_+=\partial_+ q_+^{-1} (1+\partial_+^2) + 4q_+\partial_+
\end{equation}
and by $L_+^T$ its conjugate:
\begin{equation}
	L_+^T=-(1+\partial_+^2)q_+^{-1}\partial_+-4\partial_+ q_+
\end{equation}
Eq. (\ref{LPsi}) can be written as:
\[
	L_+\psi=2
\]
For a function $f(\tau^+)$ consider the tangent 
vector $V_f$ to the phase space ${\cal M}_{SG}$ given by the equation:
\begin{equation}
	V_f q_+= -L_+^T f 
\end{equation}
Notice that ${\cal L}_{\varphi}$ is generated by vectors of the form
$V_f$ for all the possible functions $f$.
Let us  compute the commutator $[V_g, V_f]$:
\begin{eqnarray}
&&	[V_f,V_g].q_+=\nonumber\\[5pt]
&& 	
=-(1+\partial_+^2)q_+^{-1}
\left( q_+^{-1}\partial_+ f \partial_+^2 q_+^{-1} \partial_+ g
+4 \langle\partial_+ q_+ q_+ ^{-1}\rangle  g \partial_+ f -
(f\leftrightarrow g)\right)+\nonumber
\\[5pt]
&& +\left(
4\partial_+\langle 
(1+\partial_+^2) q_+^{-1}\partial_+ g + 4 \partial_+ q_+g \rangle f -
(f\leftrightarrow g)\right)=V_{[f,g]}.q_+
\end{eqnarray}
where\footnote{To avoid a confusion, we should stress that this formula
is true only if $f$ and $g$ do not depend on $\varphi$.}
\begin{equation}
[f,g]= - \langle q_+^{-1} \partial_+ f\rangle 
\stackrel{\leftrightarrow}{\partial_+}
\langle q_+^{-1}\partial_+ g\rangle
+4f\stackrel{\leftrightarrow}{\partial_+} g
\end{equation}
This verifies the Frobenius condition\footnote{The Frobenius condition
for a finite collection of vector fields on a manifold says that
if we take the commutator of any two vector fields from this
collection, this will be expressed as some linear
combination of vector fields from this collection, 
with the coefficients some functions on the manifold
(like the $\kappa$-symmetry of the superparticle).
If this condition is satisfied, we can find  ``integral manifolds''
which are tangent to these vector fields.
}
and shows that the distribution
${\cal L}_{\varphi}$ is integrable. 
This means that the sine-Gordon phase space 
is foliated by submanifolds of codimension three
such that the tangent space to the submanifold at the point $\varphi$ is
precisely ${\cal L}_{\varphi}$.
We will denote these submanifolds by the same letter ${\cal L}_{\varphi}$.
We will say more about the geometrical meaning of ${\cal L}_{\varphi}$
in the next section.

If $\xi\in {\cal L}_{\varphi}$ and $\zeta$ is tangent to the fiber
(that is, $\zeta$ changes only $\psi$ and does not affect $\varphi$) then
$\omega(\xi,\zeta)=0$. Therefore $\omega$ correctly defines a
2-form on each ${\cal L}_{\varphi}$. We will denote this 
2-form by the same letter $\omega$.
Let us evaluate $\omega(V_f, V_g)$. We have:
\[
	\omega = \delta\alpha, \;\;\;\alpha= \int d\tau^+ \psi \delta q_+
\]
\[
	\omega(V_f,V_g)=V_f.\alpha(V_g)-V_g.\alpha(V_f)-\alpha([V_f,V_g])
\]
Notice that 
\[
\alpha(V_g)=-\int d\tau^+ \langle L_+^T g \rangle \psi=
-\int d\tau^+ g L\psi =-2\int d\tau^+ g
\]
Therefore $V_f.\alpha(V_g)=0$ and we have:
\[
	\omega(V_f,V_g)=-\alpha([V_f,V_g])=
	2\int d\tau^+ [f,g]=
4\int d\tau^+ f(\partial_+ q_+^{-1}\partial_+ q_+^{-1}\partial_+ +4\partial_+) g
\]
This means that on ${\cal L}_{\varphi}$:
\begin{equation}\label{OmegaOnL}
	\omega=4\int d\tau^+ 
	\delta q_+ L_+^{-1} 
	(\partial_+ q_+^{-1}\partial_+ q_+^{-1} \partial_+
	+4\partial_+) (L_+^T)^{-1}\delta q_+
\end{equation}
Of course this form is well-defined only if both $\delta q_+$
are tangent to  ${\cal L}_{\varphi}$.
The corresponding Poisson structure $\theta=\omega^{-1}$ is a bivector
tangent to ${\cal L}_{\varphi}$:
\begin{equation}\label{BivectorTangent}
	\theta=L_+^T (\partial_+ q_+^{-1}\partial_+ q_+^{-1} \partial_+
	+4\partial_+)^{-1} L_+
\end{equation}
This formula means that the Poisson bracket between $q_+$ on the
light cone $C^+$ is:
\begin{equation}
	\{q_+(\tau_1^+), q_+(\tau_2^+)\} =
	\theta \delta(\tau_1^+-\tau_2^+)
\end{equation}
where the operator $\theta$ acting on the delta-function
on the right hand side is given by the formula (\ref{BivectorTangent})
with $q_+=q_+(\tau^+_1)$ and $\partial_+={\partial\over \partial \tau^+_1}$.
Yet another way to say it is that, given the functional $F$ on the
phase space of the sine-Gordon model, the Hamiltonian vector field 
generated by $F$ using the Poisson structure $\theta$ is:
\begin{equation}\label{HamiltonianFlow}
\dot{q}(\tau^+)=\theta{\delta F\over\delta q_+}(\tau^+)
\end{equation}
where $\delta F$ is the variational derivative\footnote{For example,
for $F=\int d\tau^+ q^2(\tau^+)$ we have 
${\delta F\over \delta q_+}(\tau^+)=2q(\tau^+)$, and for
$F=\int \cos 2\varphi$ we have 
${\delta F\over \delta q_+}(\tau^+)=2\partial_+^{-1}\sin 2\varphi(\tau^+)=
\int d\tau^+_1 \varepsilon(\tau^+-\tau_1^+)\sin 2\varphi(\tau^+)$.}
of $F$.
It is useful to compare (\ref{HamiltonianFlow}) to the Hamiltonian
vector field generated by $F$ using the standard Poisson structure
$\theta_0=\partial_+$.
The standard Poisson structure would be 
$\{q_+(\tau_1^+),q_+(\tau_2^+)\}=\delta'(\tau_1^+-\tau_2^+)$, and 
the standard Hamiltonian vector field of $F$ would be 
$\dot{q}=\partial_+ {\delta F\over\delta q_+}$, for example
$F=\int d\tau^+ \cos 2\varphi$ would generate 
$\dot{q}=2\sin 2\varphi=-4\partial_-q_+$.

Notice that Eq. (\ref{BivectorTangent}) can be written as:
\begin{equation}\label{Compatibility}
	\theta=-(\theta_1+\theta_0)\theta_1^{-1}(\theta_1+\theta_0)
\end{equation}
Here $\theta_0=\partial_+$ is the standard Poisson structure of sine-Gordon
and $\theta_1$ is the second Poisson structure\footnote{We have learned about
this second Poisson structure from \cite{Plethora}.}:
\begin{equation}
\theta_1=\partial_+^3+4\partial_+ q_+ \partial_+^{-1} q_+ \partial_+
\end{equation}
Two Poisson brackets are called compatible if their sum is again a Poisson
bracket (satisfies the Jacobi identity).
Eq. (\ref{Compatibility}) shows that $\theta$ is compatible with the standard
Poisson structure of the sine-Gordon. Indeed, 
the compatibility of two brackets $\theta_a$ and
$\theta_b$ is a bilinear condition, which we can denote 
$[\![ , ]\!]$:
\[
	[\![ \theta_a, \theta_b]\!]=0
\]
The bilinear operation $[\![ , ]\!]$ is called 
Schouten bracket.
It follows from our construction that 
$\theta=(\theta_0+\theta_1)\theta_1^{-1}(\theta_0+\theta_1)$ is
a Poisson bracket: $[\![\theta,\theta]\!]=0$ (because it corresponds to the
closed 2-form $\omega$). We have to prove that
$[\![\theta,\theta_0]\!]=0$.  We have
$[\![\theta_0,\theta_1]\!]=0$ because $\theta_0$ and $\theta_1$ are
compatible Poisson structures of the sine-Gordon model 
(we know it from \cite{Plethora}). 
Given that $\theta=\theta_1+2\theta_0+\theta_0\theta_1^{-1}\theta_0$
we have to prove that $[\![\theta_0, \theta_0\theta_1^{-1}\theta_0 ]\!]=0$.
This is true because $\theta_0^{-1}+\varepsilon\theta_1^{-1}$
is a closed 2-form for an arbitrary $\varepsilon$; therefore
$(\theta_0^{-1}+\varepsilon\theta_1^{-1})^{-1}$ is a Poisson structure
for an arbitrary $\varepsilon$; at the first order in $\varepsilon$
this means that $[\![\theta_0,\theta_0\theta_1^{-1}\theta_0]\!]=0$.

Let us rewrite $\theta$ in the following way:
\begin{equation}\label{NonlocalPB}
	\theta=-\theta_1-2\theta_0-(\partial_++4q_+\partial_+^{-1} q_+)^{-1}
\end{equation}
where
\begin{eqnarray}
&&	\theta_0=\partial_+ \nonumber\\
&&	\theta_1=\partial_+^3+
	4\partial_+ q_+\partial_+^{-1} q_+\partial_+
\end{eqnarray}
This means that although $\theta$ is nonlocal, the nonlocality
is rather weak. There are two nonlocal pieces. One is coming
from $\partial_+^{-1}$ in $\theta_1$. This can be represented by one
integration\footnote{ {\em Note in the revised version:} 
This nonlocality is related to imposing the Virasoro constraint. 
In this paper $\theta$ is the canonical Poisson bracket of the 
classical string which follows from the classical action 
$\int(\partial_+ {\bf n},\partial_- {\bf n})$ 
with the imposed Virasoro constraints $(\partial_+ {\bf n})^2=(\partial_-\bf{n})^2=1$.
If we did not impose the Virasoro constraint we would get $\theta_1$ local,
as in \cite{Bihamiltonian}.
}:
\begin{equation}
	\partial_+^{-1}f(\tau)={1\over 2}\int d\tau_1 \varepsilon(\tau-\tau_1)
	f(\tau_1)
\end{equation}
The other nonlocality comes from
\begin{equation}\label{Nonlocality}
	(\partial_++4q_+\partial_+^{-1} q_+)^{-1}=
	{1\over 2}\left({1\over \partial_++2iq_+}+{1\over \partial_+ -2iq_+}\right)
\end{equation}
The kernel of this operator also requires just one integration:
\begin{equation}
	f(\tau)\mapsto {1\over 2}\int d\tau_1  
	\varepsilon(\tau-\tau_1)\cos[2\varphi(\tau)-2\varphi(\tau_1)]
	f(\tau_1)
\end{equation}
Given a functional $F$ on the phase space, we can consider the
corresponding Hamiltonian vector field 
\[
	\dot{q}=\theta\delta F
\]
The nonlocality of the Poisson bracket leads to some ambiguities
in the definition of $\dot{q}$. One ambiguity comes from the nonlocality
in (\ref{Nonlocality}), and the other one from the $\partial^{-1}$
in $\theta_1$. Therefore $\dot{q}$ is defined up to 
$C_1\partial_+ q + C_2 \partial_- q$ where $C_1$ and $C_2$ are constants. 
This ambiguity reflects the fact that reparametrizations of the
worldsheet are gauge symmetries of the string sigma-model.
There is also a third ambiguity $\dot{q}=C_3 \cos 2\varphi$, but
we think that this vector field should probably be discarded because
of its behaviour at $\tau^+=\pm\infty$.
Also notice that the vector fields
$\dot{q}=\partial_+ q$ and $\dot{q}=\partial_-q$ are strictly
speaking not tangent to ${\cal L}_q$. Formally
$\partial_+ q= L_+^T 1$ and $\partial_- q= L_+^T \cos (2\varphi)$, but
$1$ and $\cos 2\varphi$ are not going to zero at infinity.
If this is a problem, it should be resolved by imposing
the appropriate periodicity conditions.

In some sense, the nonlocality of $\theta$ could reflect the fact
that the classical string is perhaps more sensitive to the boundary
conditions than the standard sine-Gordon model.

\section{Classical string and its dual.}
\label{sec:UsualString}
In this section we will consider the usual classical string
with the action $\int d\tau^+ d\tau^- (\partial_+{\bf n},\partial_-{\bf n})$.
With the periodic boundary conditions the
string worldsheet has a topology of the cylinder, and the
string phase space is a principal $O(3)$
bundle over the subspace of the sine-Gordon phase 
space \cite{NeveuPapanicolaou}. 
We will here choose different boundary conditions.
Let us consider the classical strings interpolating between the two
fixed lightlike geodesics. This means that in the conformal 
coordinates 
 ${\bf n}(\tau,\sigma)|_{\sigma=-\infty}$
and ${\bf n}(\tau,\sigma)|_{\sigma=+\infty}$ are 
two different equators of the sphere.
On the field theory side this corresponds to an infinite spin chain
interpolating between two different BMN vacua \cite{BMN}. 
We will call these boundary conditions the ``BMN boundary conditions''.
These boundary conditions break the $O(3)$ invariance. Therefore
we now have a map into the sine-Gordon phase space on an infinite
line, which is an injective map\footnote{A map is injective if
it is a one-to-one map  on its image.} rather than a projection.
We want to describe the symplectic form on the image of this
map which corresponds to the symplectic form of the classical
string. 
We will do it by comparing
the classical
string to its dual.

Let us consider the extended classical theory which has fields
$\lambda, \psi$ and $\phi_{\pm},A_{\pm}$ and besides that
also the field $\bf n$ with the values in $S^2$, and some
choice of the basis in the tangent space to $S^2$.
The relation between $\bf n$ and $\phi_{\pm}$ is 
\begin{equation}
	\phi_{\pm}^i=({\bf e}^i,\partial_{\pm}{\bf n})
\end{equation}
where ${\bf e}^i$ is the basis in the tangent space.
The equations of motion for the fields are 
(\ref{FirstGaugeInvariant})---(\ref{LastGaugeInvariant}).
Consider the
one-form $\hat{\alpha}$ on this extended
phase space, given by the following integral of the
local expression over the spacial contour:
\begin{eqnarray}
&&	\hat{\alpha}=\oint \left[
	(\lambda,\delta \phi_-)d\tau^- + (\lambda,\delta\phi_+)d\tau^++
	\right.\nonumber\\[5pt]
&&	+\psi \delta A_- d\tau^- +\psi \delta A_+ d\tau^+ -\nonumber\\[5pt]
&&	\left.-(\delta {\bf n},\partial_-{\bf n}) d\tau^- +
	(\delta {\bf n},\partial_+{\bf n}) d\tau^+\right]
	\label{AlphaBeforeGaugeFixing}
\end{eqnarray}
This integral does not depend on the choice of the contour.
We will denote $\widehat{\cal M}$ the phase space of this extended
model, ${\cal M}_{\widetilde{CS}}$ the phase space considered in
Section \ref{sec:Poisson} and ${\cal M}_{CS}$ the usual phase space
of the classical string parametrized by ${\bf n}(\tau,\sigma)$.
Notice that the difference of the symplectic forms $\omega_{CS}$
and $\omega_{\widetilde{CS}}$ on the extended phase space is the
differential of $\hat{\alpha}$:
\begin{equation}\label{DifferenceOfSymplecticForms}
	\omega_{\widetilde{CS}}-\omega_{CS}=\delta\hat{\alpha}
\end{equation}
If we considerd periodic boundary conditions on ${\bf n}$ then
there would be some ambiguity in restoring ${\bf n}$ from the sine-Gordon
solution, because of the global $O(3)$ symmetry. But with
the BMN boundary conditions the $O(3)$ is broken and there is
no ambiguity. 

We will now see that the 1-form $\hat{\alpha}$ actually
vanishes. 
We can describe $\hat{\alpha}$ in the following way.
Let us compare the actions of the $O(3)$ model and its dual:
\begin{eqnarray}
	S_{O(3)}&=&\int (\partial_+ {\bf n},\partial_- {\bf n})
	\label{ActionCS}\\[5pt]
	S_{\widetilde{O(3)}}&=&\int \left[
(\phi_+,(1-\psi I)\phi_-)+\right. \nonumber\\[5pt] 
&&	\left. +(\lambda, (D_+\phi_--D_-\phi_+)) +
	\psi(\partial_+ A_- -\partial_- A_+)\right]
	\label{ActionDualCS}
\end{eqnarray}
Notice that the difference of these two actions
is zero on the equations of motion:
\begin{equation}
	(S_{\widetilde{O(3)}}-S_{O(3)})_{on-shell}=0
\end{equation}
Let us consider some finite region $D$ on the worldsheet and change
the classical solution inside this region to the other classical
solution. Under the infinitesimal variation of the classical solution
inside $D$ the variation of $S_{\widetilde{O(3)}}-S_{O(3)}$ is
equal to the integral (\ref{AlphaBeforeGaugeFixing}) over the
contour $\partial D$.
But $S_{\widetilde{O(3)}}-S_{O(3)}$ is identically zero
on the classical solutions.
This shows that the integral (\ref{AlphaBeforeGaugeFixing})
over the closed contour is zero, and therefore $\hat{\alpha}$
does not depend on the choice of the contour.

Let us now impose the Virasoro constraints $r_{\pm}=1$ and 
fix the gauge $\phi_{\pm}=e^{\pm i \varphi}$.
Then the 1-form $\hat{\alpha}$ becomes:
\begin{eqnarray}
	&	\hat{\alpha}=\oint& 
	\left[ \psi \stackrel{\leftrightarrow}{\partial}_+
	\delta \varphi\; d\tau^+ 
	-\psi\stackrel{\leftrightarrow}{\partial}_-\delta\varphi\;
	d\tau^- +\right. \nonumber\\[5pt]
&&	\left.+(\delta{\bf n},\partial_+{\bf n})d\tau^+ -
	(\delta{\bf n},\partial_-{\bf n}) d\tau^-\right]
	\label{BigAlpha}
\end{eqnarray}
We have seen that the phase space ${\cal M}_{\widetilde{CS}}$ 
(the dual of the classical string)
is an affine bundle over the sine-Gordon phase space, and we denoted
${\cal L}_{\varphi}$ the foliation by the 
integral manifolds of the distribution of the 
symplectic complements of the tangent space to the fiber. 
It turns out that for the fixed BMN boundary conditions,
the image of the classical string phase space ${\cal M}_{CS}$ in the
sine-Gordon phase space is precisely one of those 
integral manifolds ${\cal L}_{\varphi}$.
To understand this, we have to explain how to lift
the tangent space to ${\cal L}_{\varphi}$ to the tangent
space of ${\cal M}_{CS}$.
The tangent space to ${\cal L}_{\varphi}$ consists of the variations
of the form:
\begin{eqnarray}\label{DeltaQpm}
&&	\delta q_+=-L^T_+ f_+,\;\;\;
	\delta q_-=-L^T_- f_-\;\;\; \\[5pt]
&&	\mbox{where }\;
	q_{\pm}=\partial_{\pm}\varphi={1\over 2}\partial_{\pm}
	\arccos(\partial_+{\bf n},\partial_-{\bf n})\;\nonumber
\end{eqnarray}
We assume that both $f_+$ and $f_-$ are rapidly decreasing at the spacial
infinity.
To find the lift $\delta{\bf n}$ it is useful to consider the 1-form 
$\hat{\alpha}$.
Let us restrict ourselves to the $C^+$ characteristic:
\begin{equation}\label{AlphaCPlus}
	\hat{\alpha}=\int_{C^+} \left[2\psi\delta q_+ d\tau^+ 
	+(\delta{\bf n},\partial_+{\bf n})d\tau^+\right]
\end{equation}
If $\delta\varphi$ satisfies (\ref{DeltaQpm}) with $f_+$ sufficiently
rapidly decreasing at $\tau^+=\pm\infty$ then:
\begin{equation}\label{AlphaFirstTerm}
	\int_{C^+} 2\psi\delta q_+ d\tau^+= 
	-4\int_{C^+} f_+d\tau^+
\end{equation}
For each tangent vector to ${\cal L}_{\varphi}$
we have to find the corresponding lift $\delta {\bf n}$ such that
$\delta \varphi =\delta\left( {1\over 2}\arccos 
(\partial_+{\bf n}, \partial_-{\bf n})\right)$ satisfies (\ref{DeltaQpm}).
The lift should be in the kernel
of $\hat{\alpha}$, because otherwise the phase space of the classical
string would have a 1-form given by the integral 
of a local expression.
Eqs. (\ref{AlphaCPlus}) and (\ref{AlphaFirstTerm})  suggests that we define
$\delta{\bf n}$ by the following equation:
\begin{equation}\label{Ansatz}
	(\delta{\bf n},\partial_{\pm}{\bf n})=4f_{\pm}
\end{equation}
This is consistent with Eq. (\ref{DeltaQpm}).
 Indeed, Eq. (\ref{Ansatz}) implies that
the variation of the field $\bf n$ is given by the following formula:
\begin{equation}\label{DeltaN}
	\delta{\bf n}=
	{4\over \sin^2 2\varphi}
	\left[(f_+-f_-\cos 2\varphi)\partial_+{\bf n} +
	(f_--f_+\cos 2\varphi)\partial_-{\bf n}\right]
\end{equation}
The equations of motion for $f_+$ and $f_-$  
follow from the constraints
$(\partial_+{\bf n},\partial_+{\bf n})=1$ and
$(\partial_-{\bf n},\partial_-{\bf n})=1$:
\begin{eqnarray}
	f_-=-{\sin 2\varphi\over 2 q_+}\partial_+ f_+ + f_+ \cos 2\varphi
	\\
	f_+=-{\sin 2\varphi\over 2 q_-}\partial_- f_- + f_-\cos 2\varphi
\end{eqnarray}
These two equations imply:
\begin{equation}
	\partial_+f_-=-{1\over 2}\sin 2\varphi \partial_+ q_+^{-1}\partial_+f_+
	-2q_+ \sin 2\varphi f_+
\end{equation}
and
\begin{equation}
	\partial_+\left({\partial_-f_+\over \sin (2\varphi)}\right)=
	-{1\over 2} q_+^{-1}\partial_+f_+
\end{equation}
This means that:
\begin{equation}
	\delta q_+ =
	-2 \partial_+\left(
	{\partial_+ f_-+\partial_- f_+\over \sin 2\varphi}\right)=
	 -L^T_+ f_+
\end{equation}
This shows the consistency of (\ref{DeltaQpm}) with 
(\ref{Ansatz}),\hspace{4pt}(\ref{DeltaN}).

We see that the 
one-form $\hat{\alpha}$ vanishes on the lift of ${\cal L}_{\varphi}$.
Therefore Eq. (\ref{DifferenceOfSymplecticForms}) implies that 
the symplectic form on ${\cal L}_{\varphi}$ following from the
classical string is given by the same
formula (\ref{OmegaOnL}) as the symplectic form following
from the dual of the classical string.
In this sense, we can say that as  classical field theories, the
theory (\ref{ActionCS}) and its dual (\ref{ActionDualCS}) are
equivalent modulo some zero modes which are not visible in the
sine-Gordon description.

The general fact of the canonicity of the nonabelian duality
was discussed 
in \cite{CurtrightZachos,KlimcikSevera,Lozano,Sfetsos}.

\section{Summary}
We considered the nonabelian dual of the classical string 
on ${\bf R}\times S^2$.
We have shown that there is a projection map from the phase space of this
model to the phase space of the sine-Gordon model. The space of functionals
on the phase space of the classical string has a subspace consisting
of the functionals of the sine-Gordon field $\varphi$. 
This subspace is closed under
the Poisson bracket of the classical string, which 
corresponds to the nonlocal Poisson
bracket (\ref{NonlocalPB}) of the sine-Gordon. 
This nonlocal Poisson bracket is compatible
with the canonical Poisson bracket of the sine-Gordon which comes from
the sine-Gordon action. (In a sense that their sum is also a Poisson
bracket.)

This suggests that the quantization (after imposing the Virasoro constraints)
of the string sigma-model on ${\bf R}\times S^2$ could be closely related
to the quantization of the sine-Gordon model with the nonlocal
Poisson structure (\ref{NonlocalPB}). 
But notice that the correspondence works only after imposing the
Virasoro constraints. From the point of view of the string theory,
it would be interesting to find a good integrable description of the
string worldsheet CFT {\em without} imposing the Virasoro
constraints. 

\section*{Acknowledgments}
I want to thank J.~Maldacena, A.~Tseytlin and K.~Zarembo for discussions.
This research was supported by the Sherman Fairchild 
Fellowship and in part
by the RFBR Grant No.  03-02-17373 and in part by the 
Russian Grant for the support of the scientific schools
NSh-1999.2003.2.


\providecommand{\href}[2]{#2}\begingroup\raggedright\endgroup

\end{document}